# Loop Quantum Theory Applied to Biology and Nonlinear Whole Biology

*Yi-Fang Chang*

*Department of Physics, Yunnan University, Kunming 650091, China*

(E-mail: yifangchang1030@hotmail.com)

**Abstract**  The loop quantum theory, which constitutes a very small discontinuous space, as new method is applied to biology. The model of protein folding and lungs is proposed. In the model, some known results are used, and four approximate conclusions are obtained: their structures are quantized, their space regions are finite, various singularities correspond to folding and crossed points, and different types of catastrophe exist. Moreover, it is discussed that the fractal is combined with a Gambini'¯s quantity. Further, based on the inseparability and correlativity of the biological systems, the nonlinear whole biology is proposed, and four basic hypotheses are formed. It may unify reductionism and holism, structuralism and functionalism. Finally, the medical meaning of the theory is discussed briefly.

Key words: loop quantum theory, biology, nonlinearity, protein folding, lungs.

PACS: 04.60.-m, 87.10+e, 04.60.Pp

**1. Introduction**

In the molecular biology the protein folding in space is a very basic question and a very complex process. It is a hotspot of protein chemistry, structure biology and molecule biology, and has been researched by many experiments and various theories [1-10]. The theoretical methods include the molecular mechanics, molecular dynamics, Monte-Carlo method, quantum mechanics, etc.

A main method of calculation is the well-known thermodynamic hypothesis proposed by Anfinsen [1]: The three-dimensional structure of a native protein in its normal physiological milieu is the one in which the Gibbs free energy of the whole system is lowest. It is that the native conformation is determined by the totality of inter-atomic interactions and hence by the amino acid sequence in a given environment. But, some experiments proved that kinetics affects protein folding and the shapes of sub-steady states, at least very important effect [3]. Moreover, Bowie, et al., investigated a method to identify protein sequences that fold into a known three-dimensional structure [2]. Chan, et al., studied transition states and folding dynamics of proteins and heteropolymers [4]. Shikhnovich, et al., discussed the transition coordinate for protein folding [5], and universally conserved positions in protein folds [7], and the protein folding nucleus using molecular dynamics [10]. Klimov, et al., researched the lattice models for proteins reveal multiple folding nuclei for nucleation-collapse mechanism [6]. Micheletti, et al., investigated that an extremality principle underlying protein evolution is shown to be possibly associated with the emergence of secondary structures [8]. Lorch, et al., discussed the effects of mutations on the enthalpy and entropy of the transition state in thermodynamics of a protein folding reaction [9].

The loop theory of the nonlinear quantum gravity is a fascinating problem in theoretical physics [11-17]. It constitutes a very small discontinuous loop space, and provides a natural embedding of the constraint surface in the phase space of Einstein theory into that of Yang-Mills gauge theory. Ashtekar introduced a complex coordinate on the extended phase space, and given a certain complexified SU(2) connection [13]. Based on the canonical quantization, Ashtekar,

Rovelli and Smolin investigated the nonperturbative quantum gravity, and these methods of loop variable have opened up bridges between gravity and other areas in mathematics and physics such knot theory, Chern-Simons theory and Yang-Mills theory [17]. Then Ashtekar, et al., introduced a black hole sector of nonperturbative canonical quantum gravity. In the theory of loop quantum gravity the fabric of space is like a weave of tiny threads, and area comes in discrete units: each thread poking through a surface gives it a little bit of area [18,19].

In this paper, the loop quantum theory is applied to biology, for instance, protein folding and the structure of lungs.

## 2. The biological model of the loop quantum theory

In the loop theory, Ashtekar, Smolin, et al., introduced new variables $\sigma_i^a$ (the square root of three-metric) and $A_a^i$ (the potential for the self-dual part of the curvature). The dynamical equations are [12]:

$$\dot{\tilde{\sigma}}^a = \sqrt{2}\,{}^{\pm}D_b \{i\underline{T}\tilde{\sigma}^{[b}\tilde{\sigma}^{a]} + T^{[b}\tilde{\sigma}^{a]}), \tag{1}$$

$${}^{\pm}\dot{A}_a = \frac{1}{\sqrt{2}}([i\underline{T}\tilde{\sigma}^{b},{}^{\pm}F_{ab}] - T^{b\pm}F_{ab}). \tag{2}$$

For the gravity including matter [15], the new equations for metric $g_{ab}$ are

$$G_{ab} + \Lambda g_{ab} = \frac{1}{\sqrt{2}}\sigma_{bAA'}[\xi^{A'}\nabla_a\xi^A - (\nabla_a\overline{\xi}^{A'})\xi^A + \overline{\eta}^A\nabla_a\eta^{A'} - (\nabla_a\overline{\eta}^A)\eta^{A'}]$$

$$-\frac{1}{4}\varepsilon_{ab}^{cd}\nabla_c k_d + 8\pi E_{ab}(KG) + 8\pi E_{ab}(YM). \tag{3}$$

Where $G_{ab}$ is the Einstein tensor, $E_{ab}(KG)$ and $E_{ab}(YM)$ are the standard stress-energy tensors of the Klein-Gordon and Yang-Mills fields. The equations of a massive Dirac spin-1/2 field $(\xi^A, \eta_{A'})$ are:

$$\sigma_{AA'}^a(\nabla_a - \frac{3i}{8}k_a)\xi^A = \frac{im}{\sqrt{2}}\eta_{A'}, \tag{4}$$

$$\sigma_{AA'}^a(\nabla_a + \frac{3i}{8}k_a)\eta^{A'} = \frac{im}{\sqrt{2}}\xi_A, \tag{5}$$

where

$$k_a = -i\sqrt{2}\sigma_a^{AA'}(\overline{\xi}_{A'}\xi_A - \overline{\eta}_A\eta_{A'}), \tag{6}$$

and $\overline{\xi}^A, \overline{\eta}^A$ are the complex conjugate variables. All equations (1)-(5) are nonlinear. It is very difficult that these equations are solved exactly. But, the loop theory of the nonlinear quantum gravity constitutes a very small discontinuous loop space, so it provides a useful method for biology.

We think that the loop quantum theory may first describe qualitatively the protein folding and the structure of lungs, in which the alveoli pulmonum corresponds to the vacuum loop. The method may be applied to describe a knot theory [17] and quantum area [18,19].

Next, the imaginary quantity number $3ik_a/8$ is deleted in Eqs. (4) and (5) one another, then we obtain:

$$\nabla_a(\frac{\eta^{A'}\xi^A}{\xi^A\xi_A + \eta^{A'}\eta_{A'}}) = \frac{im}{2\sqrt{2}\sigma_{AA'}^a}. \tag{7}$$

Assume that $\xi^A = \xi_A, \eta^{A'} = \eta_{A'}$, and let $\xi/\eta = u$, then the equation (7) becomes:

$$\frac{1-u^2}{(1+u^2)^2}\frac{du}{dt} = \frac{im}{2\sqrt{2}\sigma_{AA'}^a}. \tag{8}$$

Integral derives

$$\frac{\eta\xi}{\xi^2+\eta^2} = \frac{i}{2\sqrt{2}}\int\frac{m}{\sigma_{AA'}^a}dt + C. \tag{9}$$

In a complex plane, it is namely:

$$\sin\theta = \frac{i}{\sqrt{2}}\int\frac{m}{\sigma_{AA'}^a}dt + C. \tag{10}$$

This represents a periodic change, which is different with mass m and $\sigma_{AA'}^a$. It corresponds to a finite space region of the protein folding.

Third, for a set of the simplified differential equations:

$$x = ay + \frac{3\sqrt{2}}{8}b(x^2 - y^2)x, \tag{11}$$

$$y = ax - \frac{3\sqrt{2}}{8}b(x^2 - y^2)y, \tag{12}$$

where $\xi \to x, \eta \to y, a = im/\sqrt{2}\sigma_{AA}^a, b = \sigma_a^{AA}$. Using the method of the qualitative analysis, the characteristic matrix of these equations is:

$$\begin{bmatrix} 3\sqrt{2}b(3x^2 - y^2)/8 & a - 3\sqrt{2}bxy/4 \\ a - 3\sqrt{2}bxy/4 & -3\sqrt{2}b(x^2 - 3y^2)/8 \end{bmatrix}. \tag{13}$$

Its characteristic equation is:

$$\lambda^2 - T\lambda + D = 0, \tag{14}$$

where

$$T = 3\sqrt{2}b(x^2 + y^2)/4, D = -a^2 + 3\sqrt{2}abxy/2 - 27b^2(x^4 + y^4)/32 +$$
$$27(bxy)^2/16, \Delta = T^2 - 4D = 9b^2(x^4 + y^4)/2 + 4a^2 - 6\sqrt{2}abxy - 9(bxy)^2/2. \tag{15}$$

When D>0, there are nodal points for $\Delta$ >0; there are focal points for $\Delta$ <0. While the focal points are unstable sources for T>0, and are stable sinks for T<0. When D<0, there are saddle points. These singular points correspond to folding and cross points. For (0,0), T=0, $D = -a^2 < 0, \Delta = 4a > 0$, it is a saddle point.

Fourth, using the adiabatic approximation of synergetics in Eqs. (11) and (12), let y'=0, then

$$y^3 - x^2 y + Ay = 0, \tag{16}$$

where $A = (8ia/3\sqrt{2}b)$. The solutions of this equation of third order are:

$$y_1 = v + w, y_2 = \varepsilon_1 v + \varepsilon_2 w, y_3 = \varepsilon_2 v + \varepsilon_1 w, \tag{17}$$

where
$$v = [\frac{Ax}{2}(-1 + \sqrt{1 - \frac{4x^4}{27A^2}})]^{1/3}, \tag{18}$$

$$w = [\frac{Ax}{2}(-1 - \sqrt{1 - \frac{4x^4}{27A^2}})]^{1/3}, \tag{19}$$

$$\varepsilon_{1,2} = -\frac{1}{2} \pm i\frac{\sqrt{3}}{2}. \tag{20}$$

A solution $y_1$ is substituted for equation (11), so

$$x' = ia(v+w) + \frac{3\sqrt{2}}{8}b[x^3 - x(v^2 + 2vw + w^2)]. \quad (21)$$

It is very complex. But, let $x'=$, and based on the parallel equations (4) and (5) and (21), from the Thom's catastrophe theory we may derive a type of the folding catastrophe with single parameter:

$$V = x^3 + ax. \quad (22)$$

While there is a catastrophe of the elliptic umbilic point with single parameter:

$$V = x^3 - 3y^2 + a(x^2 + y^2) - bx - cy, \quad (23)$$

and a catastrophe of the hyperbolic umbilic point with double parameters:

$$V = x^3 + y^3 + axy - bx - cy. \quad (24)$$

In the loop space representation of quantum general relativity [16], the wave functions depend on a piecewise differentiable loop and vanish on intersecting curve, thus non-trivial contributions may arise only at the points where the loop is non-differentiable and has "corner", and there has Fig.1. When $t^-, t^+$ are the unit tangent vectors at the point $z_j$, and $t^- = t^+$, it corresponds to self-intersecting loops whose component loops are tangent at the contact point (Fig.2). This case forms two joint loops. It and two focal points correspond to lungs, and sink and source correspond to the breathing function of lungs.

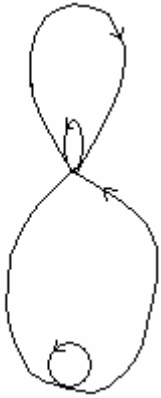
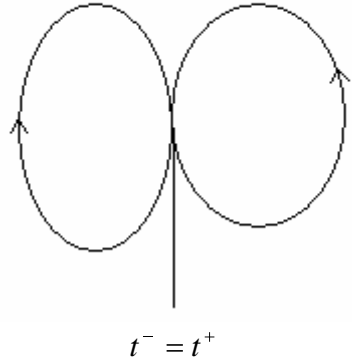

$$t^- = t^+$$

Fig.1        Fig.2

In 1989, Gambini, et al., [14] applied quantitatively a quantity

$$K = d(MassGap)/d\sqrt{\mu}, \quad (25)$$

where $\mu$ is the coupling parameter, and K=2.28, etc. In particle physics the energy gap possesses the self-similarity, for example, intermittency of the multiple productions. Such K may correspond to the fractal dimension D. D can describe many self-similarity of the protein folding. Moreover, if the lungs connect a single interacting triplet of waves, for whose strange attractor, D=2.32 [20], or connect the Lorenz model (D=2.07), it will agree with D=2.17 of lungs. Moreover, the shape of the lungs is very similar to the Lorenz butterfly, i.e., strange attractor like two cycles.

This paper combines the loop quantum theory of gravitation, and proposes a new method of protein folding and lungs. But, it is consistent in essence with the basic idea of the general relativity: The matter determines the structure of space.

## 3. Nonlinear whole biology and its basic hypotheses

At present, the main research method of biology is continually restored to the original state or structure, from a living object to its organs, tissues, cells and biological macromolecules. In certain aspect, it is a linear reductionism of one-to-one correspondence. Schr?dinger points out that one essential character of life is its ability to show cooperative behaviors. Instead of the incoherent fluctuations of atoms or small molecules in solution, living cells show coherent global dynamics. Cooperativity has also been found to be a very important feature that can deeply affect the behavior of nonlinear systems.

In 1989, Peyrard and Bishop investigated the statistical mechanics of a simple nonlinear lattice model for the denaturation of the DNA double helix[21]. This PM model consists of two chains connected by Morse potentials representing the H bonds. Then Peyrard, et al., studied effective breather trapping mechanism for DNA transcription[22], and generation of high-energy localized vibrational modes in nonlinear Klein-Gordon lattices[23]. Further Peyrard proposed an experiment using micro-mechanical stretching of DNA to probe nonlinear energy localization in a lattice[24]. In these thermalized lattices the properties of "¨discrete breather"¡±re exact solutions of these nonlinear lattices. In biological molecules such as DNA, where large amplitude nonlinear motions are essential for function, temporary deviations from energy equipartition could play an important role. Using numerical simulations and kinetics calculations he estimates the order of magnitude of the expected force fluctuations. They examine the behavior of two types of models, which describe the melting of double-stranded DNA chains. Type-I model (with displacement-independent stiffness constants and a Morse on-site potential) is probably the simplest, exactly solvable, one-dimensional lattice model with a true thermodynamic phase transition; its scaling behavior near the critical point can be characterized by the exponents Type-II model (with displacement-dependent stiffness constants) is analyzed numerically and shown to have a first-order transition with finite melting entropy, discontinuous fraction of bound pairs, divergent correlation lengths, and critical exponents[29].

Peyrard, et al., investigated the vector nonlinear Klein-Gordon lattices, where the envelope soliton solutions of a helicoidal DNA model described by a radial and an angular degree of freedom for each site[25,26]. They discuss nonlinear dynamics of DNA, its statistical mechanics, and one of the experiments that one can now perform at the level of a single molecule and which leads to a non-equilibrium transition at the molecular scale. After a review of experimental facts about DNA, further they introduce simple models of the molecule and show how they lead to nonlinear localization phenomena that could describe some of the experimental observations. Then they analyze the thermal denaturation of DNA, and discuss some aspects of the initiation of transcription, in particular the formation of the open complex and the activation mechanism associated to enhancer binding proteins, and discuss the mechanical opening of the DNA double helix[27,28].

Recent, Peyrard, et al., studied the hydration water of proteins and its biological activity, and found a behavior typical of a proton glass, with a glass transition of about 268K. In order to analyze these results, they investigate the statistical mechanics and dynamics of a model of the hydrogen bonding scheme of bounded water molecules, and discuss the connection between the dynamics of bound water and charge transport on the protein surface as observed in the dielectric measurements[30]. Then they proves quantitatively that few nonlinear oscillators can show chaotic dynamics, and a nonlinear lattice made of such oscillators coupled to each other, and may on the contrary exhibit coherent excitations such as solitons or nonlinear localized modes. They research the fundamental properties of nonlinear lattices and their applications in condensed matter and biomolecular physics[31].

Based on the most basic features whole and nonlinearity of the biology and combining the general nonlinear theory, we discussed briefly the nonlinear whole biology [32,33]. The fundamental thought is that based on the biological structure and holism. Further, in the linear theory of any biological systems we introduce various nonlinear terms, which represent interactions in a system and among the system and other systems, and consider the circumstance factors as the boundary or initial conditions. From this, combining various known theories, the research in various respects of biology may spread out in different levels. We propose the four basic hypotheses of the nonlinear whole biology:

First hypothesis: The inseparability exists always among different parts and different levels in various biological systems, which determinates to the biological globality.

Second hypothesis: Many main characteristics, for example, self-organization, self-adjustment and self-reproduction, of biological systems are produced from some especial structures of complex subsystems. From this the interaction and nonlinearity exist necessarily.

Third hypothesis: From a biological macromolecule to a gigantic ecological system on the Earth, various biological systems of different levels possess the totality and nonlinearity.

Fourth hypothesis: A basic property of any biological systems as an open system is this system and its environment must be a whole. It corresponds to a generalized metabolism. Usually environment is regarded as a boundary condition of the system, but it and the biological systems have often various nonlinear relations.

The totality and the nonlinearity are two basic biological characters. They are closely related. Because of complexity, the inseparability, and the correlativity of the biological systems, their description must apply the nonlinear theory with the interaction terms. Every biological system is a paragon, which unified completely structure and function.

The whole appears not only in a synthesis from a lower level to higher level, but also in a unification of structure and function on biology. Reversibly, if there is not the totality, one and one molecule cannot become a living thing, one and one organ separated is not man.

The nonlinear whole biology is described totally by the nonlinear mathematics. For the nonlinear equations, we may sum up two large types:

1. An equation possesses the self-interaction or other nonlinear terms. It includes: In the quantum biology, scientists introduced the nonlinear Schr?dinger equation,

$$i\psi_t + J\psi_{xx} + G\psi^3 - A\psi = 0, \qquad (26)$$

which can describe the soliton model of the vibrational energy transported along the biological macromolecules [34,35]. Peyrard investigated the nonlinear Klein-Gordon equation of DNA [25,26]. The brain science is one of the frontiers of the modern biology. For the brain model, scientists introduced the nonlinear Duffing equation,

$$x'' + kx' + x + ex^3 = Fcon\Omega t, \qquad (27)$$

which reflects coupling and oscillation of the nervous networks, and may depict the resonance and morphogenesis. In the blood circulation system, scientists introduced the nonlinear hydrodynamic equation [36], for example, the Navier-Stokes equation. A simplified form of differential equation of second order is:

$$x'' + \alpha x' + f(x) = 0, \qquad (28)$$

where f(x) includes some nonlinear terms, e.g., $bx^3$, etc.

2. A set of coupled equations. A simplified form of two elements is:

$$x' = a_1 x + b_1 y + F_1(x, y),$$
$$y' = a_2 x + b_2 y + F_2(x, y). \tag{29}$$

Here F(x,y) are some coupled terms on x and y. It includes: In the molecular biology, the movement of DNA double helices may be described by a set of coupled sine-Gordon equations [37]. In the nonlinear enzyme dynamics, the catalytic reaction has the well-known three-molecular model, i.e., the nonlinear Brusselator. In the cell biology, scientists introduced two-cell coupled nonlinear chemical oscillator [38]. In the population dynamics there are various nonlinear models, for example, the equations of Lotka-Volterra model [39]. Kauffman [40,41] has modeled genetic regulatory network, which is a dynamical system that specifies for each such combination, or state of gene activities. The analogue to the potential wells of spin glasses is basin of attraction and attractors in these network models. Moreover, scientists discussed the molecular basis for interaction of the protein [42], the structure of TATA-box-binding protein and that of the transcription factor [43,44]. Their mathematical descriptions should be nonlinear theories. The other nonlinear theories are Jocob-Monod operon model of protein synthesis, the synergetic equations among various biological systems, and the morphogenesis in the genetics, etc. Further, all of various self-organizations in biology should correspond to the nonlinear theory. All evolutions, from biology to genes, are the nonlinear phenomena.

Since Eigen proposed the hypercycle in 1971, there is a scientific theory on the relationship between protein and nucleic acid and on the origin of life. Eigen pointed out[45]: Biological complexity appears already at the level of macromolecular chemistry, and it has always been recognized as a typical requisite of biological organization. In the hypercycle cooperative behaviour is reflected by intrinsically nonlinear reaction mechanisms, and the dynamics is described by a system of coupled nonlinear differential equations. In the hypercycle each cycle as a whole has self-enhancing growth properties, and different sets of the nonlinear equations carries information and function, etc. The hypercycle is namely a typical and very perfect theory of the nonlinear whole biology.

The nonlinear whole biology is consistent with the systems biology[46-48]. From the nonlinear whole biology view, we can unify reductionism and holism, and unify structuralism and functionalism on biology, can create and open various regions of the nonlinear whole biology, and make the description and some predictions. Moreover, the complex biological systems provide possibly some modes on decrease of entropy in the isolated system [49-51].

**4. Gauge theory and the nonlinear whole medicine.**

It is known that gravitational interaction possesses Einstein's $GL(4,R)$ symmetry of general coordinates and Weyl's $SL(2,C)$ symmetry of gauge invariance[52]. The equation of the Yang-Mills gauge theory is:
$$D_v F_a^{\mu v} = J_a^{\mu}, \tag{30}$$
where $F_a^{\mu v} = \partial^\mu A_a^v - \partial^v A_a^\mu + g C_{abc} A_b^\mu A_c^v$, $C_{abc}$ are different structure constants of various gauge groups. This equation has derived various solutions: for instance, monopole solution, dyon solution, instanton solution, meron solution, string solution, vortex solution and dilaton solution, in which some solutions possess possibly biological meaning. Different solutions correspond to various phase transformations, which are probably different bifurcations and folding.

According to the loop quantum theory and the equations of gauge field, the protein folding originates possibly from some external forces, boundary conditions, interactions, and the current in the Yang-Mills equations and their nonlinear self-interaction terms in the structural constant.

As an example, we discuss the nonlinear whole medicine. It includes the nonlinear mathematical model of the humoral immunity, etc. For the respiratory system, a usual way of quantitative description is the rheology [36]. In recent years, scientists proposed that the structure of lungs is one of fractals, and it already is nonlinear.

The diseases of the respiratory system are considered usually by two causes: 1. The lungs are invaded by the maleficent substance from the out side; 2. The diseases of other organs, for instance, heart, liver, and kidney, etc., pass through the blood vessel-lymph systems to spread to lungs. In the loop theory of the nonlinear whole medicine, the external invasions correspond to the external forces, and to the boundary conditions; the internal infections correspond to the interaction terms, e.g., the current in the Yang-Mills equations. In this case, the nonlinear theory points out another result, the nonlinear self-interaction terms, which relates to the structural constant of the gauge theory, and to $k_a$ in Eqs. (4) (5), corresponds to a type of diseases derived from the smaller substratum, for instance, cell, gene, etc. They include cancer, sarcoidosis, and diffuse interstitial fibrosis of lungs. The occurrence of these diseases has randomness from the view of macroscopic respiratory system, and the internal and external infections of body are only the external conditions of this type of diseases. The above two types of diseases have usually necessary, for example, all people cannot be avoided under poison gas. For third disease, the equation (28) may be simplified to a well-known nonlinear equation form:

$$X' = (aX + b)X, \qquad (31)$$

whose bifurcation-chaos agrees qualitatively with the cancerous change.

From the view of the nonlinear whole biology, the above classification of diseases possesses universality. The medical treatment ways of different diseases should possess some different characteristics. For example, the external infections of body should separate from the maleficent substance; the diseases of internal infections make that the traditional medical treatments possess rationality; the decrease of the randomness of occurrence on third diseases must investigate the smaller substratum. Further, it will benefit by the research on difference and relation between the modern and traditional medicine.